\documentclass[a4paper,11pt]{article}
\usepackage{jinstpub} % for details on the use of the package, please see the JINST-author-manual
\usepackage{lineno}
\title{\boldmath Retrieval-Augmented Question Answering over Scientific Literature for the Electron-Ion Collider}

\ifx
\author{Tina. J. Jat, T. Ghosh, Karthik Suresh}
\affiliation{Ramaiah University of Applied Sciences, India}
\affiliation{College of William \& Mary, USA}
\fi

\author{Tina. J. Jat$^{1}$, T. Ghosh$^{1*}$, Karthik Suresh$^{2}$}
\affiliation{$^{1}$ Ramaiah University of Applied Sciences, India}
\affiliation{$^{2}$ College of William \& Mary, USA}

\emailAdd{tapasi03@gmail.com}

\abstract{To harness the power of Language Models in answering domain specific specialized technical questions, Retrieval Augmented Generation (RAG) is been used widely. 
In this work, we have developed a Q\&A application inspired by the Retrieval Augmented Generation (RAG), which is comprised of an in-house database indexed on the arXiv articles related to the Electron-Ion Collider (EIC) experiment - one of the largest international scientific collaboration and incorporated an open-source LLaMA model for answer generation. This is an extension to it's proceeding application built on proprietary model and Cloud-hosted external knowledge-base for the EIC experiment. This locally-deployed RAG-system offers a cost-effective, resource-constraint alternative solution to build a RAG-assisted Q\&A application on answering domain-specific queries in the field of experimental nuclear physics. This set-up facilitates data-privacy, avoids sending any pre-publication scientific data and information to public domain. 
Future improvement will expand the knowledge base to encompass heterogeneous EIC-related publications and reports and upgrade the application pipeline orchestration to the LangGraph framework.
}

\keywords{Electron-Ion Collider (EIC), Retrieval-Augmented Generation (RAG), Large Language model Meta AI (LLaMA), Retrieval-Augmented Generation Assessment(RAGAS)}

\begin{document}
\maketitle
\flushbottom

\ifx
\section{Planning}
Intro and organization of the paper - 1 page
previous work and literature survey - 1 page
objective of our work- 1/2 page
methodology, schematic and workflow-1.5 page
code and packages used - 1.5 page
Result - 1.5 page
Summary-1/2 page
future and challenges - 1/2 page
\fi

\section{Introduction}
\label{sec:intro}
%Language, the primary medium of communication thoroughout the evolution of human history, plays an important role in any societal advancements. 
In the era of Artificial Intelligence(AI), Natural Language Processing(NLP) and Language Models(LM) are the domains that encode the essence of the meaning of language based on its surrounding or context to facilitate AI-driven reasoning and decision making tasks. Any advancement in this domain has created an enormous implications to the human lives across domains including science, technology, commerce and so on.
As a consequence, any bias or confident sounding factually incorrect generated outcome imputed in those applications will create a domino effects across domains among the various stakeholders. The present state-of-the art Language Models either proprietary or open-source which are trained on exorbitantly large amount of text available in the internet exhibit this problem of generating fluent, confident but factually incorrect responses popularly known as hallucination~\cite{maynez2020faithfulness, ji2023survey}. In 2021, Meta research team has introduced the concept of Retrieval Augmented Generation aka RAG~\cite{lewis2021rag} to mitigate hallucination and make the LMs grounded.
In our research work we have implemented this methodology to build a Question-Answering application to generate reliable, scientifically accurate answer for the Electron-Ion Collider(EIC)~\cite{abdul2022eic} experiment.
EIC being a multi-continental, multi-institutional projects usually produce large amount of scholarly articles in the form of research publication, conference/workshop presentation, working group meeting presentations, technical design report etc. 
A large scale international collaboration with more than 190 participating institutes across the world provides an excellent opportunity to deploy such an AI-based application to facilitate smooth, time-efficient on-boarding to the newly joined collaborators as well as the seasoned researchers to extract vital scientific and key research objectives/milestones of the experiment along with detailed description of user-specific questions from various scientific details of the theory, simulation, detectors, hardware specification catering the needs of the user.

With this ambition the AI4EIC~\cite{suresh2024rag} built the foundational work on RAG Q\&A system with proprietary models and Cloud-based storage for external knowledge-base. In this present work, we have further extended this application using cost-effective open-source Language Models.% which is capable of executing and deploying the application in compute constraint situation. 
Another major advancement is the use of in-house database to index the EIC-related articles extracted from arXiv repository.
The generated answers are traced back to the original arXiv articles from which the context is being retrieved. Emphasis is given on the authenticity of this citation mechanism which is facilitated through LangSmith~\cite{langsmith}, an observability platform for LLM applications, allowing any interested reader to dive deeper into the subject through those cited references.  
%The performance of the RAG-system is evaluated with the standard RAG-assessment(RAGAS)[ref] metrics.

% organization 
This paper is organized in the following way: a brief review of the previous work is presented in Section~\ref{sec:prev_work}, the methodology adopted is detailed in Section~\ref{sec:method}, results and evaluation performance are reported in Section~\ref{sec:result} and finally conclusions along with a brief road-map for future work are provided in Section~\ref{sec:summary}.

%%%%%---------------------------------------------------------------------------------------------------------%%%%%
\section{Previous Work}
\label{sec:prev_work}
The present development on the RAG-based Q\&A application for the EIC project has broadened the foundational work~\cite{suresh2024rag}.
In the said application, a benchmark Q\&A dataset with gold standard is being created using the OpenAI GPT4.0 model.
The dataset contains specific sets of 51 questions, and for each question there is associated sub-questions referred as "claims",
answers against each claim and a comprehensive response.
These questions are referenced from EIC-related articles from arXiv repository published after 2021.
More details about the strategy and the quality of this dataset can be found in~\cite{suresh2024rag}. The pipeline also facilitates generation of new Q\&A annotations by user through a web-based chatbot interface.
In this work, the authors used OpenAI’s \texttt{text-embedding-ada-002} model to embed context and Pinecone~\cite{pinecone} to store the embeddings into a knowledge base.
OpenAI GPT4.5 model generated the answer.
%The retrieved documents are wrapped into a prompt template and fed into an OpenAI GPT4.5 model.
The entire RAG pipeline is orchestrated by LangChain framework and traced by the LangSmith, including the citations of resources.
%They have reported slightly lower context relevancy and context accuracy owing to the number of retrieved text using either Cosine similarity or MMR techniques.

%%%%%---------------------------------------------------------------------------------------------------------%%%%%
\ifx
\section{Methodology and Application Architecture}
%\label{sec:method}
As mentioned the novelty of this work is on building an in-house knowledge base and for that we have explored several tools e,g., Faiss~\cite{douze2024faiss}, ChromaDB~\cite{chromadb_site} however ChromaDB is chosen for being more user friendly and resource efficient.
We reused the chunking philosophy of the earlier work~\cite{suresh2024rag}, however additionally cross-validated with varying chunk sizes e.g. 120 and 180 with overlapping context of 20 characters between two consecutive chunks.
Chunking refers to the technique of splitting large text documents into smaller segments called "chunks"~\cite{yepez2024financial}.
The chunks are embedded using the \texttt{mxbai-embed-large} by Mixedbread AI~\cite{mxbai2024embed} embedding model of 1024 dimensional vector.
chunk size is an important factor in elevating the performance of a RAG-based application, since smaller chunks may increase precision but risk missing essential information, while larger chunks capture more content but may compromise relevance~\cite{jiang2024longrag}.
% (fine balance while selecting chunking method and chunking size) (mention the chunking strategy employed here).

%For better inference during later stage of retrieval, the metadata such as arXiv id, authors, year of publication etc, are also embedded and stored along with the chunked data in the VectorDB.
For better inference during the later stages, several metadata associated with each article such as arXiv ID, authors, publication year etc, are merged with chunked text, embedded and stored in the vector DB. This additional information helps the model differentiate between semantically similar chunks from different sources, improving contextual accuracy and traceability. This approach allows the system to refer to the correct citations rather than giving isolated or ambiguous text snippets.

%----Q. WHy ChromaDB over Pinecone and pros and cons
Several database options were considered, including FAISS~\cite{douze2024faiss}, Pinecone~\cite{pinecone}, LanceDB~\cite{lancedb}, and ChromaDB~\cite{chromadb_site}, to efficiently store the embedded chunks. While FAISS offers high-performance similarity search and Pinecone provides a managed, cloud-based service with scalability, ChromaDB was chosen due to its combination of practical advantages of data privacy.
%Being open-source and freely available, it allows full control over data and infrastructure without any additional cost (cost here is compared to the pinecone cloud usage cost as it has limitations on the free version.). 
ChromaDB supports local deployment, seamless integration into frameworks such as LangChain, simplifying the construction of RAG pipelines. The architecture of the end-to-end RAG is demonstrated in Figure~\ref{fig:qa-system-design}.

%Note: Need to add 2 line on query to vector and the package used:
When a user submits a query, it is encoded using the \texttt{mxbai-embed-large} embedding model, consistent with the embeddings of the documents stored in the vector database. A similarity search is then performed between the encoded query and the stored embeddings to retrieve the most relevant document.

%---- Retrieval: Cosine similarity and MMR
The retrieval stage seeks to retrieve documents that best match the intent of the user's query. The effectiveness of the retrieved documents heavily depends on both the quality of the embeddings and the strategy used to measure similarity, as they directly influence which documents are considered relevant (pertinent). Although many similarity metrics have been proposed in prior research, this study highlights cosine similarity and Maximum Marginal Relevance (MMR) as the primary methods for retrieving documents.
Cosine Similarity is the fundamental metric for retrieving chunks most relevant to a user's query, measuring the alignment between two non-zero embedding vectors by the cosine of the angle between them~\cite{steck2024}. By emphasizing vector orientation, it captures semantic similarity - where 1 indicates perfect alignment, 0 no similarity, and -1 semantic opposition. However, it can produce redundant results and ignores vector magnitude, which may encode information such as importance or confidence~\cite{zhou2022}. MMR extends this by balancing relevance and diversity: after an initial cosine-based retrieval, it iteratively selects documents that are both relevant and minimally similar to previous chosen ones~\cite{goldstein1998}. This reduces the redundancy and improves the diversity of the retrieved content~\cite{juseondo2025}.
\fi

%%%%%---------------------------------------------------------------------------------------------------------%%%%%
%----------------rewriting the methodology-------------------------
\section{Methodology and Application Architecture}
\label{sec:method}

\begin{figure}[htbp]
\centering
\includegraphics[width=.8\textwidth]{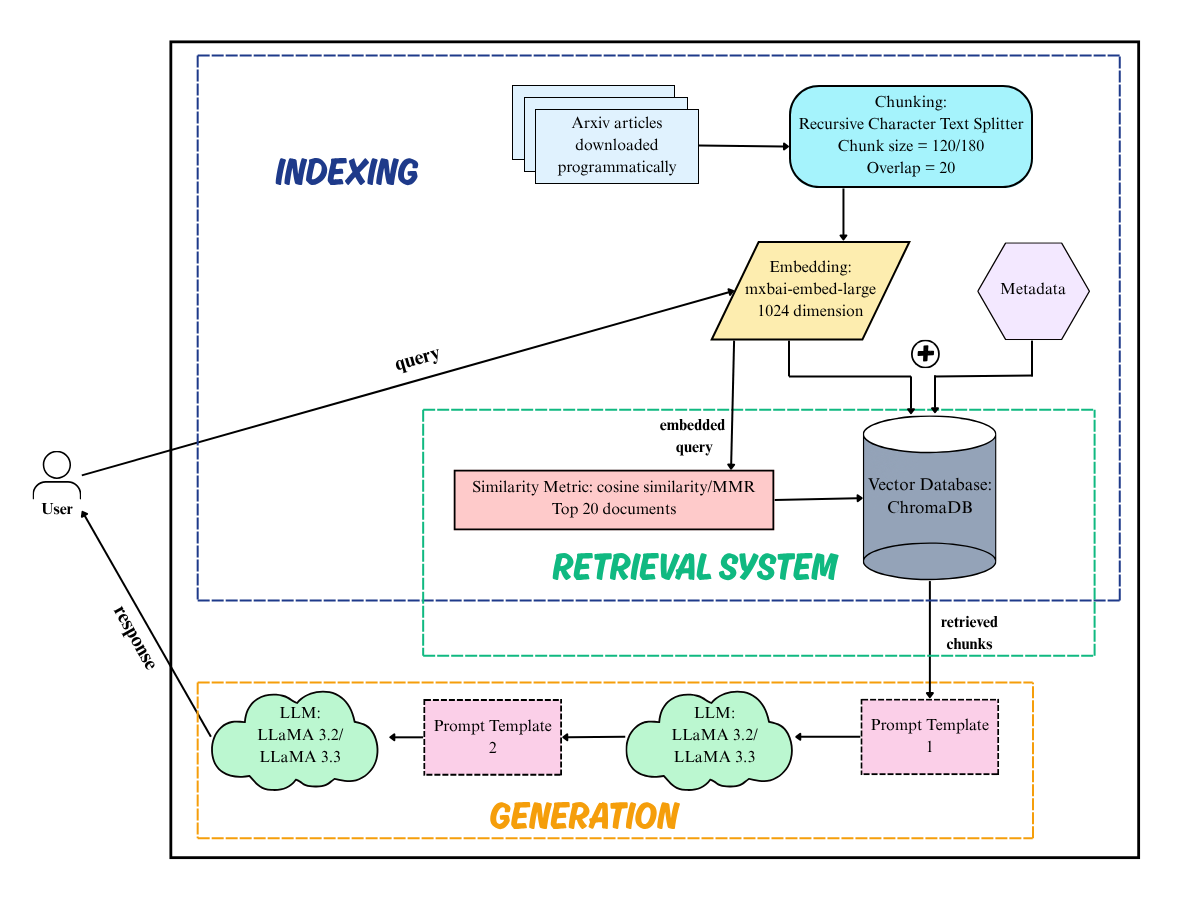}
\caption{The schematic of the Q\&A System Design for the EIC. The pipeline consists of article ingestion, document chunking, vector embedding, ChromaDB indexing, retrieval, formation of prompting to propagate the information and response generation. First, a query is encoded into a dense-vector representation during the data ingestion phase; subsequently similarity search is initiated between the embedded query and locally deployed vectorized database to retrieve the most relevant contexts. These retrieved chunks are subsequently merged with the query through a carefully designed prompt template and passed to a language model, which finally generates a contextually grounded response.\label{fig:qa-system-design}}
\end{figure}

The Q\&A application implements a RAG-pipeline consisting of five major steps: context ingestion and pre-processing, chunking, embedding generation, retrieval via similarity search and LLM-conditioned answer generation encoded through prompt template. The schematic of the architecture is demonstrated in Figure~\ref{fig:qa-system-design}.

%corpus
The knowledge base of this RAG-inspired Q\&A application is constructed with 178 EIC-related research articles from the arXiv preprint repository.
These scholarly articles span research domains across phenomenology, software development, detector design, accelerator physics etc. To enhance inference and retrieval fidelity, several metadata associated with each article such as arXiv ID, authors, year of publication etc, are concatenated with chunked text, then embedded and stored in the vector database. This additional information helps the model differentiate between semantically similar chunks from different sources, improving contextual accuracy and traceability. This approach allows the system to refer to the correct citations rather than giving isolated or ambiguous text snippets.

% chunking
The selected arXiv articles are divided into fixed-sized smaller text segments, known as "chunks"~\cite{yepez2024financial}.
%This chunking~\cite{yepez2024financial} refers to the technique of splitting large text documents into smaller segments called "chunks".
\texttt{RecursiveCharacterTextSplitter} from LangChain~\cite{langchain_splitter} splits each document into "chunks" of lengths 120 and 180 characters with a 20 character overlap between consecutive chunks. This overlap preserves semantic continuity across chunks and mitigate any artificial fragmentation during inference.
We followed the chunking philosophy of the earlier work~\cite{suresh2024rag}, however additionally cross-validated with chunk sizes e.g. 120 and 180 characters.
Determining the optimal chunk size is an important factor in elevating the performance of a RAG-based application, since smaller chunks may increase precision but risk missing essential information, while larger chunks capture more content but may compromise relevance~\cite{jiang2024longrag}.

%Embedding
Subsequently, each of these chunks are embedded into a 1024 dimensional dense vector representation using the \texttt{mxbai-embed-large} model provided by Mixedbread AI~\cite{mxbai2024embed}. \texttt{mxbai-embed-large} is a transformer-based model that facilitates local deployability, has a strong performance record in the Massive Text Embedding Benchmark(MTEB)~\cite{mteb2022} and no API-dependency. % vector DB
The external knowledge base is saved into a persistence storage system. Several database options were explored, including FAISS~\cite{douze2024faiss}, Pinecone~\cite{pinecone}, LanceDB~\cite{lancedb}, and ChromaDB~\cite{chromadb_site}, to efficiently store the embedded chunks. While FAISS offers high-performance similarity search; Pinecone provides a managed, cloud-based service with scalability, ChromaDB was chosen due to its combination of practical advantages of data privacy through local deployment and seamless integration into the LangChain orchestration framework, simplifying the construction and implementation of the RAG pipeline.
%and native integration with the LangChain orchestration framework.

%retrieval strategy
The retrieval stage extract the most semantically aligned documents related to the user's query. The users' query is encoded into a 1024-dimensional vector using the \texttt{mxbai-embed-large} embedding model, the same dimension as the indexed documents stored in the vector database.
%The semantic similarity between the encoded query and the stored embeddings are captured either by Cosine Similarity or MMR approach.
The retrieval stage seeks to retrieve documents that best match the intent of the user's query. The effectiveness of the retrieved documents heavily depends on both the quality of the embeddings and the strategy adopted to encode the semantic similarity, as they directly influence the most pertinent documents. Although many similarity metrics have been proposed in prior research, this study adopted two strategies: cosine similarity and Maximum Marginal Relevance(MMR)~\cite{goldstein1998}.

Cosine Similarity measures the angular alignment between two non-zero embedding vectors by the cosine of the angle between them~\cite{steck2024}. The angle between them captures semantic similarity and the most similar retrieved contexts attains the top scores. However, it is invariant to the magnitude of the embedding vectors which may encode information such as importance or confidence, enabling to produce redundant results~\cite{zhou2022}. MMR extends this by balancing relevance and diversity: after an initial cosine-based retrieval, it iteratively selects documents that are both relevant and minimally similar to previous chosen ones~\cite{goldstein1998}, which reduces the redundancy and improves the diversity of the retrieved content~\cite{juseondo2025}.

% answer generation and citation tracing
Top 20 retrieved chunks are concatenated into a prompt and passed through either LLaMA 3.2 or LLaMA 3.3 model, which are deployed on-premises without any API-dependency. The prompt provides guardrails to restrict answer generation grounded on the provided context only and refrain from any erroneous responses. A citation tracing mechanism is orchestrated through LangSmith~\cite{langsmith}, which traces the generated answers to specific arXiv articles from which the relevant and supporting context is retrieved, thereby grounding the answers. LangSmith facilitates the recording and tracing of each step of the inference pipeline: the user query, retrieved content along with the associated metadata, the prompt which propagated these information to the LMs and the generated answer. This allows the Q\&A system system into a transparent research assistance. 

%Evaluation
Finally the generated answers are evaluated by the RAG assessment metrics and the details are elaborated in~\ref{sec:result}.

%%%%%---------------------------------------------------------------------------------------------------------%%%%%
\section{Results Analysis and Discussion}
\label{sec:result}
To infer of the performance of the RAG-system, the generated answers are evaluated by comparing with an expert curated dataset, a collection of the EIC-related questions and their optimal solutions.
%-----
\subsection{Benchmark Dataset}
The AI4EIC2023\_DATASETS is the high quality benchmark dataset that contains the ground truth answers of a set of 51 questions~\cite{suresh2024rag}. 
To evaluate the output generated from the RAG-application, this gold standard dataset is curated. These ground truth question-answers are generated using GPT-4.0 model, contextualized from the EIC-related publications from the arXiv pre-print repository across domains e.g., high energy phenomenology(hep.ph), nuclear experiments(nucl.ex) etc. Each question in this dataset is mapped to a pre-defined number of sub-parts called "claims", individual answer against each claim and a comprehensive answer of the entire question.
The AI-generated QA-pairs are meticulously vetted by human experts to create a gold-truth for validating the RAG-generated responses.
There is also provision to generate question-answers pairs by an annotator directly from user-chosen arXiv articles following the similar Q\&A structure of "claims", supporting responses etc. from the web-based annotation interface.

%------
\ifx
This study leverages AI4EIC2023\_DATASETS benchmark dataset, which was specifically
designed for evaluating EIC related RAG based QA system by using an LLM-assisted approach
for creating high-quality dataset, as detailed in [1]. This novel dataset was meticulously created
using GPT-4.0, which processed EIC related publications from multiple groups(eg. theory,
simulation, experiment, software development) from arXiv disciplines, ranging from hep.ph to
nucler to ph-acc2. AI generated QA pairs are created from those articles. Each question in the
dataset is explicitly linked to a defined number of "claims" and the corresponding answer that
specifies individual claim, ideal responses for each, and a comprehensive overall response.
There is a provision to generate QA pairs from users query directly from the web-based
annotation interface, which was designed to enable human annotators to select arXiv papers,
define the number of questions and claims, and revise/edit the LLM produced QA pairs. This
AI-assisted process ensures that the LLM generated responses are grounded to specific arXiv
articles, making the dataset highly suitable for assessing the retrieval and reasoning abilities of
RAG system.
\fi

%%%%%---------------------------------------------------------------------------------------------------------%%%%%
\subsection{Evaluation and result analysis}
\begin{figure}[htbp]
\centering
\includegraphics[width=0.8\textwidth]{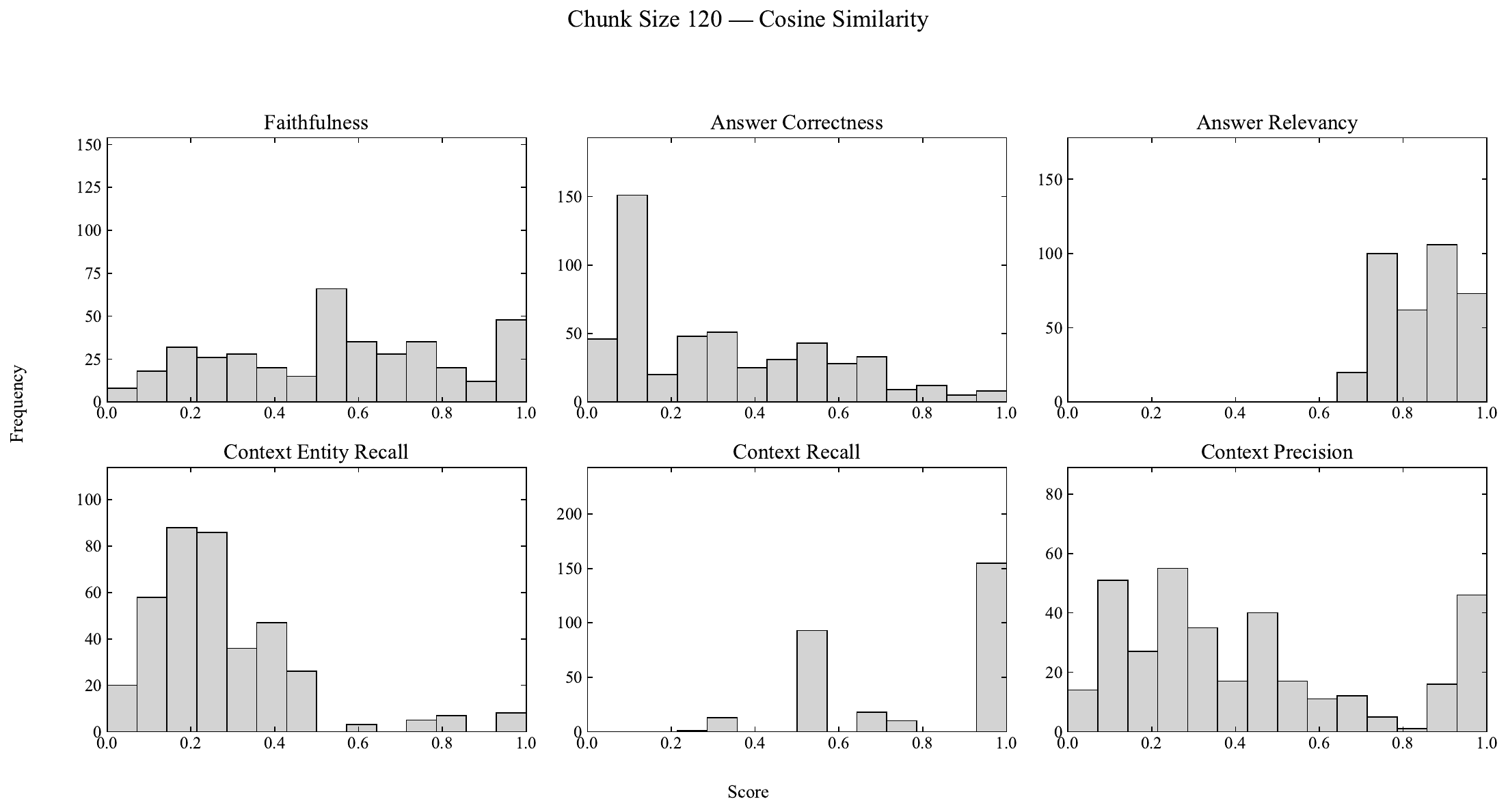}
\vspace{0.5cm} % optional vertical space between images
\includegraphics[width=0.8\textwidth]{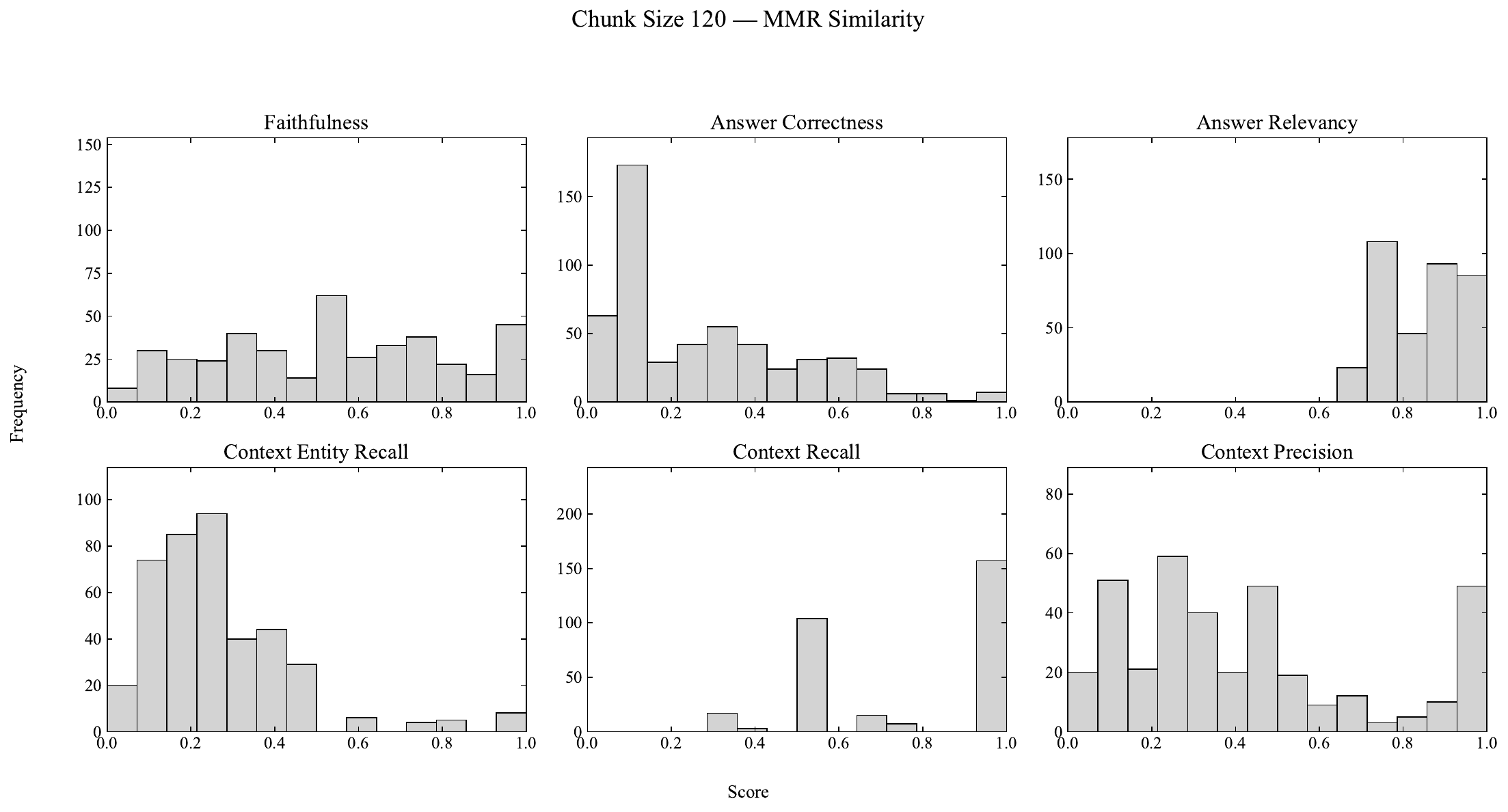}
\caption{The performance of the Q\&A is represented by the six RAGAS evaluation metrics for retrieval and answer generation. Upper panel: The evaluation metrics for chunk size of 120 characters for Cosine similarity; Lower panel: Same metrics as measured for MMR retrieval technique and 120 chunk size.}
\label{fig:120-chunks}
\end{figure}

\begin{figure}[htbp]
\centering
\includegraphics[width=0.8\textwidth]{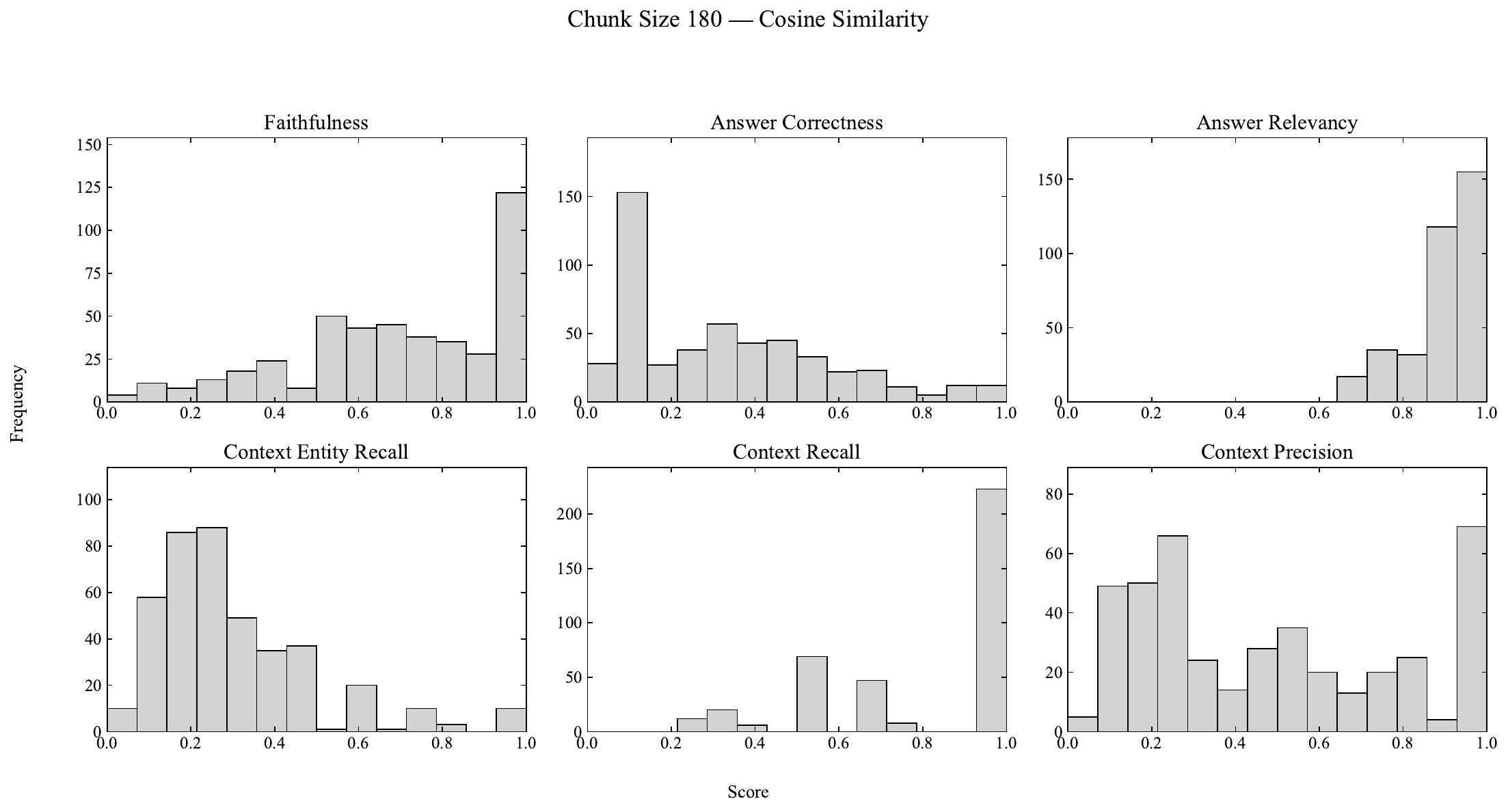}
\vspace{0.5cm} % optional vertical space between images
\includegraphics[width=0.8\textwidth]{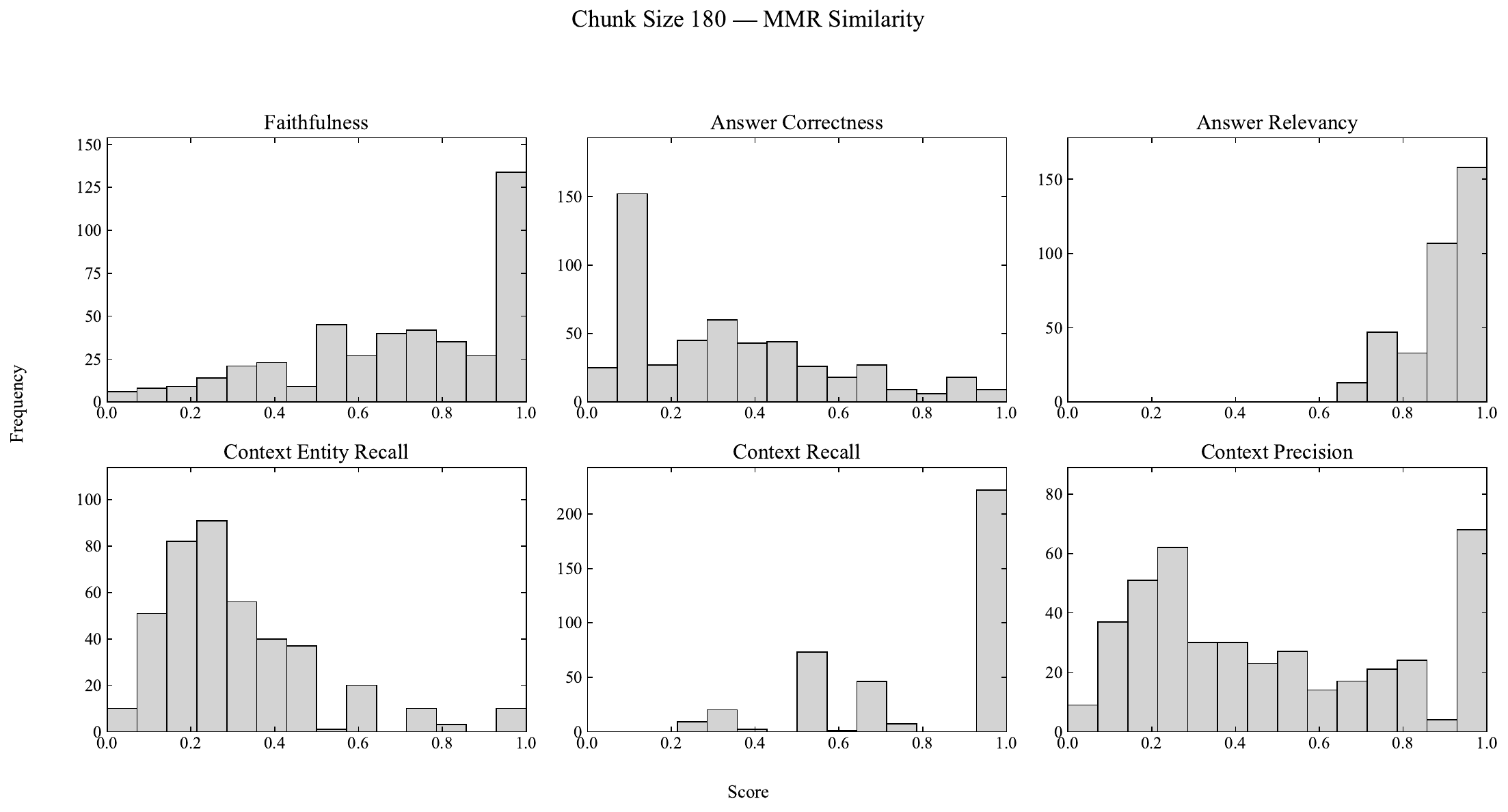}
\caption{The generation performance as characterized by the six metrics of RAGAS framework for chunk size of 180 characters with Cosine similarity(upper panel) and MMR retrieval(lower panel) respectively.\label{fig:180-chunks}}
\end{figure}

The performance of the application is evaluated with the RAGAS framework~\cite{es2025ragas} encoded into a set of 6 evaluation metrics; Context Entity Recall, Context Precision, Context Recall, Answer Relevancy, Answer Correctness and Faithfulness.
The first three metrics are to validate the groundedness and factuality of the generated answers with respect to the retrieved context, where as the later three metrics encode the factual accuracy and semantic similarity of the generated response. We have explored the evaluation scores across four different configurations: chunk sizes 120 and 180 and similarity metrics cosine \& MMR for context retrieval. The distributions scores for all these combinations are shown in Figure~\ref{fig:120-chunks} and Figure~\ref{fig:180-chunks}.
%--------- 
%%3. Evluation - latency retrieval and inference
The performance of the end-to-end RAG-pipeline is evaluated by measuring latency of the  retrieval and answer generation steps independently. The retrieval systems' performance is encapsulated via the retrieval latency: the time elapsed between the query submission and the return of the most relevant 20 documents retrieved from the knowledge base.

\begin{figure}[htbp]
\centering
\includegraphics[width=.4\textwidth]{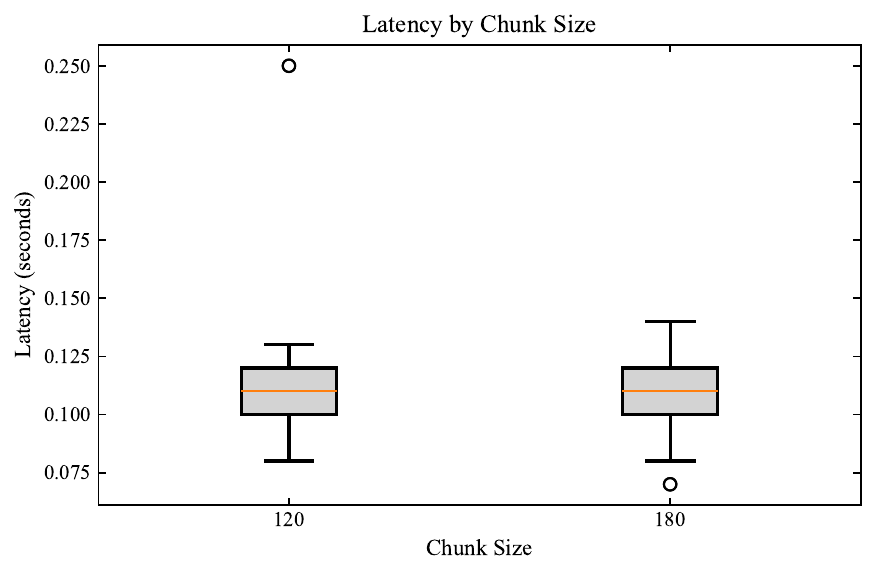}
\qquad
\includegraphics[width=.4\textwidth]{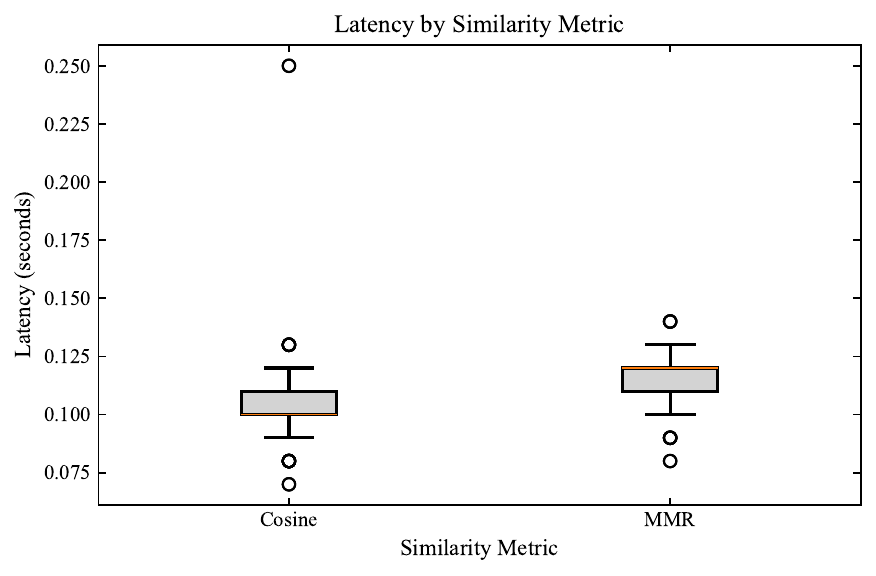}
\caption{Left : Retrieval latency of the RAG-system for chunk size 120 and 180 characters with overlap of 20 characters between two consecutive chunks. Right: Retrieval latency measured for two two different similarity retrieval mechanism: Cosine similarity and MMR.\label{fig:lat-chunk}}
\end{figure}

% latency interpretation
Figure~\ref{fig:lat-chunk} shows the retrieval latency distributions for chunk sizes - 120 and 180, and for similarity metrics - Cosine and MMR. It is represented as box plots to highlight both the central tendency and the variability. The median latency for chunk size of 120 characters is 0.11s, whereas for 180 it is 0.11-0.12s.
It is also observed that the choices of similarity metrics and the chunk size yield similar latency period highlighting no significant advantage of one over the other.
However, the larger chunk introduces slightly wider variability.

\begin{figure}[htbp]
\centering
\includegraphics[width=.4\textwidth]{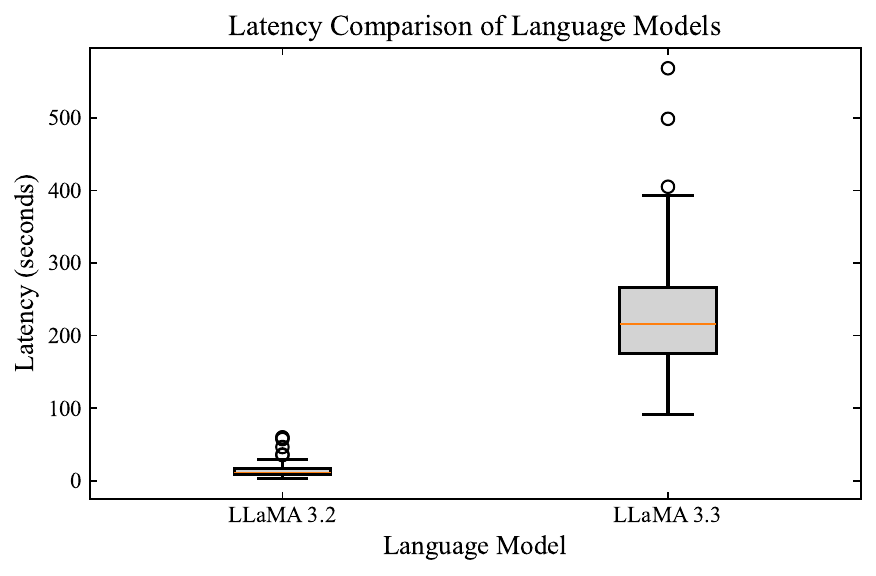}
\caption{Latency distribution during answer generation of the RAG-application for two different Language Models LLaMA 3.2 and LLaMA 3.3 respectively.\label{fig:lat-llm}}
\end{figure}

%interpretation
On the contrary, the choice of LLMs has a significant impact on the inference latency: the time taken for token generation by the LMs after prompt submission.
The most prominent variation is observed in the Figure~\ref{fig:lat-llm}, where LLaMA 3.2 outperforms LLaMA 3.3 model performance by an order of magnitude. The LLaMA 3.3 model utilizes more compute and exhibits substantially higher and more varying latency.
LLaMA 3.2 depicts stability with median latency of 10-20 sec, a narrow inter-quartile range and moderate outliers of $~$50-60 sec.
Whereas LLaMA 3.3 yields a drastic shift in median latency. It shows wider variability accompanied by extreme outliers.
This higher computation overhead of the larger model is not suitable for a Q\&A chatbot application and hence we incorporated the LLaMA 3.2 model for further study.

%-----------about the metrices
The RAG pipeline is evaluated with the RAGAS metrics to access both retrieval and answer generation steps.

%% 1. Evaluation - 120 chunk size

%% for retrieval
The performance of the retrieval system is captured through three metrics namely: Context Entity Recall, Context Precision and Context Recall. Context Entity recall~\cite{ragas_context_entities_recall} examines the proportion of the entities from the ground truth that are been recalled in the retrieved context. Whereas Context Recall~\cite{ragas_context_recall} captures the percentage of the number of claims in the ground truth that have been extracted in the retrieved context. The Context Precision~\cite{ragas_context_precision} reveals the retrieved chunks that are relevant to the user query, hence the proportion expresses the precision of the retrieved chunks. 
%Among these three metrics Context Recall shows robust performance across combinations with scores clustered around 1.0 highlighting that the significant proportions of claims of the ground truth are extracted in the retrieved context. The performance is improved for chunk size 180 where frequencies are increased near to score 1.0. This improvement is aligned with the hypothesis that the longer chunk preserve the coherence and semantic of the text. The Context Precision exhibits moderate bimodal distribution with scores spread across both low end $[0.1,~0.3]$ and high values above 0.9. Context Precision distribution is broadly invariant among similarity Cosine angle and MMS. The Context Entity Recall metric had wide spread across low to moderate scores. 
Among these three metrics Context Recall shows robust performance across combinations with scores clustered around 1.0 highlighting that the significant proportions of claims of the ground truth are extracted in the retrieved context. The performance is improved for chunk size 180 where frequencies are increased near to score 1.0. This improvement is aligned with the hypothesis that the larger chunk preserve the coherence and semantics of the text. The Context Precision exhibits moderate bimodal distribution with scores spread across both low end within $[0.1,~0.3]$ and high values above 0.8. Context Precision distribution is broadly invariant among similarity choices and the chunk lengths. The Context Entity Recall metric distribution is wide, spread across low to moderate scores for all the four combinations. This under-performance highlights the limitation of the retrieval mechanism to extract scientific named entities as the dense embedding models are optimized for general semantic not for any specific scientific terminology.  
%interpretation - retrieval(above)

%interpretation - Answer generation
As mentioned earlier the quality of the generated answer is encoded through three RAGAS evaluation metrics.
%% for answer generation
Answer Relevance measures semantic alignment between the generated answer and the
query by calculating the mean of the cosine similarity between the original question and a set of reverse-engineered questions derived from the generated answer~\cite{ragas_answer_relevance}.
Similarly, the Answer Correctness~\cite{ragas_answer_correctness} is the weighted average of the semantic similarity and factual consistency of the answer. The factual consistency is encoded through the F1-score while the cosine angle between the ground truth answer and the generated answer captures the semantic relevance. The Faithfulness metric measures the number of "claims" in the generated answer which are also supported by the claims in the retrieved context~\cite{ragas_faithfulness}, which aids in identifying and quantifying the hallucination.
As shown in Fig~\ref{fig:120-chunks}, Faithfulness score exhibits wider variability for 120 chunk, whereas for 180 chunk the distribution is strongly right-skewed and achieves more than 90\% scores for large number of instances. This demonstrates that larger chunk size renders richer contextual representation and more factual responses, as the proportions of claims in retrieved context are usually higher than the smaller chunk size irrespective of the choice of retrieval strategy. 
A similar trend is also reflected in the Answer Relevancy score. For 180 chunk size, the distribution is concentrated largely above $0.9$ score where as for 120 chuck size it is bimodal and wide ranged. 
The Answer Correctness encapsulates both the factual similarity and the groundedness of the generated answer with respective to the ground truth. The scores are poor across all the combinations, which is likely due to the EIC-experiment specific complex factual details and also the lightweight LLaMA 3.2 language model.

%% 2. Evaluation - 180 chunk size

%-----------------------
\section{Conclusion}
\label{sec:summary}
% general
This work presents proof-of-concept design, implementation and quantitative evaluation of a RAG-based Question-answering system tailored to publications related to the EIC-experiment. The system is deployable on-premises and built on open-source components which provide a cost-effective and secure alternative to Cloud-based storage and proprietary model. The initiative is aligned towards data primacy and operational independence of large international scientific collaboration.
The pipeline is constructed with an external knowledge base consisting of PDF articles published in arXiv, open-source components: the \texttt{mxbai-embed-large} model for text embedding, ChromaDB for persistent storage, LLaMA models for answer generation. The pipeline is orchestrated through LangChain, while citation tracing is enabled by LangSmith.
%from latency
LLaMA 3.3 results an order-of magnitude increase in inference latency along with wider variability and extreme outliers for some queries. This larger model may lead to better reasoning ability, however this could not be verified in this work owing to the compute constraint.
The result also highlights that the model scaling has greater impact on the latency than the design choices such as chunk size and similarity metrics.
The RAGAS evaluation established that chunk-size 180 offers optimal configuration; however, in our study, MMR mechanism does not demonstrate any added advantage over cosine similarity. % inspite of offering reduced redundancy among the retrieved context. 
% Future direction - different modalities, LangGraph, Agentic RAG with MCP
Future work will focus on incorporating additional resources, such as PowerPoint presentations, wiki, white paper, reports, etc. into the knowledge base. An imminent major planned upgrade of the pipeline is to migrate to LangGraph~\cite{langgraph} orchestration framework.
%-------------------------
\ifx
For internal references use label-refs: see section~\ref{sec:intro}.
Bibliographic citations can be done with "cite": refs.~\cite{a,b,c}.
When possible, align equations on the equal sign. The package
\texttt{amsmath} is already loaded. See \eqref{eq:x}.
\begin{equation}
\label{eq:x}
\begin{aligned}
x &= 1 \,,
\qquad
y = 2 \,,
\\
z &= 3 \,.
\end{aligned}
\end{equation}
Also, watch out for the punctuation at the end of the equations.

If you want some equations without the tag (number), please use the available
starred-environments. For example:
\begin{equation*}
x = 1
\end{equation*}

\section{Figures and tables}

All figures and tables should be referenced in the text and should be
placed on the page where they are first cited or in
subsequent pages. Positioning them in the source file
after the paragraph where you first reference them usually yield good
results. See figure~\ref{fig:i} and table~\ref{tab:i} for layout examples. 
Please note that a caption is mandatory and its position differs, i.e.\ bottom for figures and top for tables.

\begin{figure}[htbp]
\centering
\includegraphics[width=.4\textwidth]{example-image-a}
\qquad
\includegraphics[width=.4\textwidth]{example-image-b}
\caption{Always give a caption.\label{fig:i}}
\end{figure}

\begin{table}[htbp]
\centering
\caption{We prefer to have top and bottom borders around the tables.\label{tab:i}}
\smallskip
\begin{tabular}{lr|c}
\hline
x&y&x and y\\
\hline
a & b & a and b\\
1 & 2 & 1 and 2\\
$\alpha$ & $\beta$ & $\alpha$ and $\beta$\\
\hline
\end{tabular}
\end{table}

We discourage the use of inline figures (e.g. \texttt{wrapfigure}), as they may be
difficult to position if the page layout changes.

We suggest not to abbreviate: ``section'', ``appendix'', ``figure''
and ``table'', but ``eq.'' and ``ref.'' are welcome. Also, please do
not use \texttt{\textbackslash emph} or \texttt{\textbackslash it} for
latin abbreviaitons: i.e., et al., e.g., vs., etc.

\paragraph{Up to paragraphs.} We find that having more levels usually
reduces the clarity of the article. Also, we strongly discourage the
use of non-numbered sections (e.g.~\texttt{\textbackslash
  subsubsection*}).  Please also consider the use of
``\texttt{\textbackslash texorpdfstring\{\}\{\}}'' to avoid warnings
from the \texttt{hyperref} package when you have math in the section titles.

\appendix
\section{Some title}
Please always give a title also for appendices.

\fi

\acknowledgments

T. Ghosh acknowledges the research support received from Ramaiah University of Applied Sciences though out the preparation of this work.

%\paragraph{Note added.} This is also a good position for notes added after the paper has been written.

% Bibliography

%% [A] Recommended: using JHEP.bst file
%% \bibliographystyle{JHEP}
%% \bibliography{biblio.bib}

%% or
%% [B] Manual formatting (see below)
%% (i) We suggest to always provide author, title and journal data or doi:
%% in short all the informations that clearly identify a document.
%% (ii) please avoid comments such as "For a review'', "For some examples",
%% "and references therein" or move them in the text. In general, please leave only references in the bibliography and move all
%% accessory text in footnotes.
%% (iii) Also, please have only one work for each \bibitem.

%%\bibliographystyle{IEEEtran}
%%\bibliography{biblio}

\begin{thebibliography}{99}
\ifx
\bibitem{a}
Author,
\emph{Title},
\emph{J. Abbrev.} {\bf vol} (year) pg.

\bibitem{b}
Author,
\emph{Title},
arxiv:1234.5678.

\bibitem{c}
Author,
\emph{Title},
Publisher (year).
\fi

\bibitem{maynez2020faithfulness}
J. Maynez, S. Narayan, B. Bohnet and R. McDonald,
\emph{On faithfulness and factuality in abstractive summarization},
in \emph{Proc. 58th Annual Meeting of the Association for Computational Linguistics (ACL)}, 2020, pp. 1906--1919,
arXiv:2005.00661.

\bibitem{ji2023survey}
Z. Ji, N. Lee, R. Frieske et al.,
\emph{Survey of hallucination in natural language generation},
\emph{ACM Comput. Surv.} \textbf{55} (2023) 1,
arXiv:2202.03629.

\bibitem{lewis2021rag}
P.~Lewis et al.,
\emph{Retrieval-Augmented Generation for Knowledge-Intensive NLP Tasks},
arXiv:2005.11401.

\bibitem{abdul2022eic}
R. Abdul Khalek et al.,
\emph{Science requirements and detector concepts for the electron-ion collider},
\emph{Nucl. Phys. A} \textbf{1026} (2022) 122447,
arXiv:2103.05419.

\bibitem{suresh2024rag}
K.~Suresh, N.~Kackar, L.~Schleck and C.~Fanelli,
\emph{Towards a RAG-based Summarization Agent for the Electron-Ion Collider},
arXiv:2403.15729.

\bibitem{langsmith}
LangChain AI,
\emph{LangSmith SDK: Platform for LLM application monitoring and evaluation},
GitHub repository (2026).

\bibitem{pinecone}
Pinecone Systems Inc.,
\emph{Pinecone vector database},
online service (2026).

\bibitem{yepez2024financial}
A.~J.~Yepes, Y.~You, J.~Milczek, S.~Laverde, and R.~Li,
\emph{Financial Report Chunking for Effective Retrieval-Augmented Generation},
arXiv:2402.05131.

\bibitem{langchain_splitter}
LangChain,
\emph{RecursiveCharacterTextSplitter documentation},
online documentation (2026).

\bibitem{jiang2024longrag}
Z.~Jiang, X.~Ma, and W.~Chen,
\emph{LongRAG: Enhancing Retrieval-Augmented Generation with Long-Context LLMs},
arXiv:2406.15319.

\bibitem{mxbai2024embed}
S.~Lee, A.~Shakir, D.~Koenig and J.~Lipp,
\emph{Open Source Strikes Bread — New Fluffy Embedding Model},
Mixedbread Blog (March 8, 2024).

\bibitem{mteb2022}
N.~Muennighoff, N.~Tazi, L.~Magne and N.~Reimers,
\emph{MTEB: Massive Text Embedding Benchmark},
arXiv:2210.07316.

\bibitem{douze2024faiss}
M.~Douze et al.,
\emph{The Faiss library},
arXiv:2401.08281.

\bibitem{lancedb}
LanceDB,
\emph{LanceDB: Developer-friendly OSS embedded retrieval library for multimodal AI},
GitHub repository (2026).

\bibitem{chromadb_site}
Chroma Developers,
\emph{Chroma vector database — fast, scalable search for AI applications},
https://www.trychroma.com/.

\bibitem{goldstein1998}
J.~Goldstein and J.~G.~Carbonell,
\emph{Summarization: (1) using MMR for diversity‑based reranking and (2) evaluating summaries},
\emph{TIPSTER Text Program Phase III Workshop} (1998) 181.
doi:10.3115/1119089.1119120.

\bibitem{steck2024}
H.~Steck, C.~Ekanadham and N.~Kallus,
\emph{Is cosine-similarity of embeddings really about similarity?},
\emph{Companion Proc. ACM Web Conf.} {\bf 2024} (2024) 887,
doi:10.1145/3589335.3651526.

\bibitem{zhou2022}
K.~Zhou, K.~Ethayarajh, D.~Card and D.~Jurafsky,
\emph{Problems with cosine as a measure of embedding similarity for high frequency words},
arXiv:2205.05092.

\bibitem{juseondo2025}
J.~D.~Hwang, J.~Kwon, H.~Kamigaito and M.~Okumura,
\emph{Considering length diversity in retrieval-augmented summarization},
arXiv:2503.09249.

\bibitem{es2025ragas}
S.~Es, J.~James, L.~Espinosa-Anke and S.~Schockaert,
\emph{Ragas: Automated Evaluation of Retrieval Augmented Generation},
arXiv:2309.15217.

\bibitem{ragas_answer_relevance}
Ragas,
\emph{Answer Relevance},
Ragas Documentation (v0.1.21),
\url{https://docs.ragas.io/en/v0.1.21/concepts/metrics/answer_relevance.html}.

\bibitem{ragas_answer_correctness}
Ragas,
\emph{Answer Correctness},
Ragas Documentation (v0.1.21),
\url{https://docs.ragas.io/en/v0.1.21/concepts/metrics/answer_correctness.html}.

\bibitem{ragas_faithfulness}
Ragas,
\emph{Faithfulness},
Ragas Documentation (v0.1.21),
\url{https://docs.ragas.io/en/v0.1.21/concepts/metrics/faithfulness.html}.

\bibitem{ragas_context_entities_recall}
Ragas,
\emph{Context Entities Recall},
Ragas Documentation (v0.1.21),
\url{https://docs.ragas.io/en/v0.1.21/concepts/metrics/context_entities_recall.html}.

\bibitem{ragas_context_recall}
Ragas,
\emph{Context Recall},
Ragas Documentation (v0.1.21),
\url{https://docs.ragas.io/en/v0.1.21/concepts/metrics/context_recall.html}.

\bibitem{ragas_context_precision}
Ragas,
\emph{Context Precision},
Ragas Documentation (v0.1.21),
\url{https://docs.ragas.io/en/v0.1.21/concepts/metrics/context_precision.html}.

\bibitem{langgraph}
LangGraph,
\emph{LangGraph: Agent Orchestration Framework for Reliable AI Agents},
[online]. Available: \url{https://www.langchain.com/langgraph}, 2026.

\end{thebibliography}
\end{document}